\documentclass[trackchanges]{aastex701}

\begin{document}

\title{ Spectroscopic Study of the Circularly Polarised Radio Source TYC 2834-1385-1: An Active Binary} 

\author[orcid=0009-0009-6244-7573,sname=Bleeker, gname=Tamara Ade Safitri]{Tamara A. S. Bleeker}
\affiliation{Anton Pannekoek Institute for Astronomy, University of Amsterdam, 1098 XH Amsterdam, The Netherlands}
\email[show]{tamara.bleeker@student.uva.nl}  

\author[orcid=0009-0001-2095-3760,sname=Hoogerbrug, gname=Alexander]{Alexander Hoogerbrug} 
\affiliation{Anton Pannekoek Institute for Astronomy, University of Amsterdam, 1098 XH Amsterdam, The Netherlands}
\email[show]{alexander.hoogerbrug@student.uva.nl}

\author[orcid=0009-0007-0740-0954,gname=Elise,sname=Koo]{Elise Koo}
\affiliation{Anton Pannekoek Institute for Astronomy, University of Amsterdam, 1098 XH Amsterdam, The Netherlands}
\affiliation{ASTRON, Netherlands Institute for Radio Astronomy, Oude Hoogeveensedijk 4, Dwingeloo 7991 PD, The Netherlands}
\email{e.j.m.koo@uva.nl}

\author[orcid=0000-0002-3516-2152,gname=Rudy ,sname=Wijnands]{Rudy Wijnands}
\affiliation{Anton Pannekoek Institute for Astronomy, University of Amsterdam, 1098 XH Amsterdam, The Netherlands}
\email{R.A.D.Wijnands@uva.nl}

\author[orcid=0000-0002-7167-1819,sname=Callingham,gname=Joseph]{Joseph R. Callingham}
\affiliation{Anton Pannekoek Institute for Astronomy, University of Amsterdam, 1098 XH Amsterdam, The Netherlands}
\affiliation{ASTRON, Netherlands Institute for Radio Astronomy, Oude Hoogeveensedijk 4, Dwingeloo 7991 PD, The Netherlands}
\email{j.r.callingham@uva.nl, callingham@astron.nl}


\begin{abstract}

Close binaries with active chromospheres have recently been shown to emit circularly polarised radio emission at low frequencies, motivating detailed studies of their orbital properties. We present high-resolution ($R=85,000$) optical spectroscopy of TYC 2834-1385-1, a previously unclassified radio source, obtained with the HERMES spectrograph on the Mercator Telescope. We identify the source as a double-lined spectroscopic binary and measure radial velocities from both metallic absorption and Ca II emission lines. We find an orbital period of $P=3.601\pm0.002$ days and nearly equal radial-velocity semi-amplitudes for the two components. The system exhibits strong Ca II H and K emission, which is consistent with a BY Dra-type binary containing two similar late-G/early-K main-sequence stars, with a preliminary classification of K0V. This represents one of the first spectroscopic confirmations of a chromospherically active binary following its identification in a low-frequency radio survey.



\end{abstract}




\section{Introduction}

The circularly polarised LOFAR Two-Metre Sky Survey (V-LoTSS) has revealed bright, highly circularly polarised radio emission from chromospherically active close binaries \citep{Callingham2023}. Circular polarisation exceeding 10\% is rare among radio sources and is characteristic of coherent radio emission mechanisms \citep{2020NatAs...4..577V}, while the exact mechanism responsible remains poorly understood.


Among the classes of chromospherically active close binaries, BY Draconis (BY Dra) systems are characterised by short orbital periods ($\lesssim$10-12 days), late-type components, and strong Ca II H and K emission from their active chromospheres \citep{1976ASSL...60..287H}. Spectroscopic characterization of radio-selected sources can therefore establish whether they belong to this class and provide the orbital constraints needed to interpret their radio emission.

TYC 2834-1385-1 was identified in V-LoTSS as a circularly polarised radio source \citep{Callingham2023}, but has no previous spectroscopic observations. Its position in a Hertzsprung–Russell diagram is consistent with a late-type dwarf \citep{Callingham2023}, but its multiplicity and spectral classification were unknown. In this Research Note, we present the first high-resolution spectroscopic observations of TYC 2834-1385-1, determine its orbital parameters, and provide a preliminary spectral classification.

\begin{figure*}[ht!]
\plotone{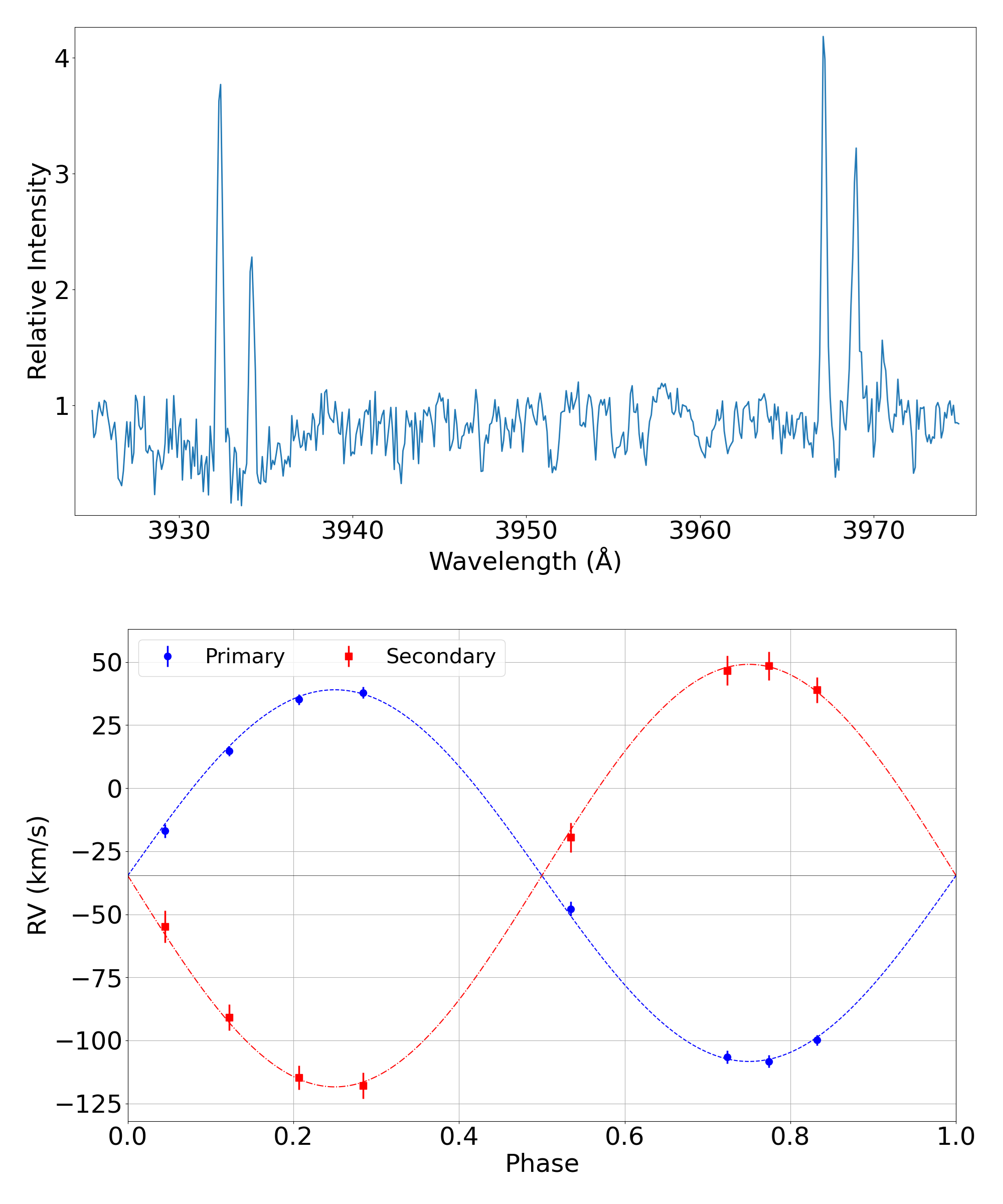}
\caption{\textbf{Top:} The Ca II H and K region of the normalised spectrum of TYC 2834-1385-1 obtained on October 6, 2025. The double-peaked emission in both the H (right) and K (left) lines is characteristic of active chromospheres in both stars. The stronger emission peaks correspond to the primary stellar component.
\textbf{Bottom:} Phase-folded radial-velocity fit of TYC 2834-1385-1. The plotted data points are derived from fits to the average metal-line profile. The corresponding Ca II H and K measurements are consistent with these values within their uncertainties.}


\label{fig:general}
\end{figure*}


\section{Observations} \label{sec:observations}

We obtained high-resolution optical spectra with the HERMES spectrograph mounted on the 1.2-metre optical Mercator Telescope \citep{raskin2011hermes}. HERMES has a spectral resolution of $R\approx85,000$ and provides wavelength coverage from approximately 3770 \AA\, to 9000 \AA. Our observations were carried out between October 6 and October 22, 2025. TYC 2834-1385-1 was observed on eight nights in 2025 (October 6, 7, 11, 12, 13, 14, 20, and 22), sampling different orbital phases to increase the likelihood of detecting radial velocity variations caused by a possible binary companion.

Each spectrum had an exposure time of 1 hour, except for the observations on October 20 and 22, which consisted of two co-added 25-minute exposures, resulting in a total integration time of 50 minutes on those nights. The spectra have a typical signal-to-noise ratio of $\sim58$ at $\lambda=500$ nm.


The raw data were processed using the standard HERMES data reduction pipeline \citep{raskin2011hermes}, which performs bias subtraction and flat-field correction using internal halogen-lamp exposures, wavelength calibration using thorium-argon and neon lamp exposures, and cosmic-ray mitigation using a standard sigma-clipping procedure.



\section{Methods and Results} \label{sec:methods}

\subsection{Radial velocity fitting} \label{subsec:rvfit}

All spectra of TYC 2834-1385-1 were normalised using a spline-fitting algorithm. The resulting spectrum shows many metallic lines and is characteristic of a relatively cool spectroscopic binary, consistent with G- or K-type spectral classifications. We employed a least-squares deconvolution method \citep{2013A&A...560A..37T} to calculate the average metallic absorption line profile across the 4397 $\AA$ to 5801 $\AA$ wavelength range for both stars, where the spectra have the highest signal-to-noise ratio. The mean stellar absorption profiles were then fitted using a double Gaussian to obtain the radial velocities and fit uncertainties for both components in each spectrum.


In addition to the abundance of metallic lines, the spectrum of TYC 2834-1385-1 shows double Ca II H and K emission lines, as shown in the top panel of Figure \ref{fig:general}. These are indicative of active chromospheres in both components of the binary \citep{1990clst.book.....J}, and are consistent with the spectra of BY Dra variables \citep{1976ASSL...60..287H}. These calcium emission lines provide additional radial-velocity measurements of the system. The double-peaked Ca II emission lines were similarly fitted with double Gaussian profiles to determine the radial velocities of their respective components.


The resulting radial-velocity measurements were analysed using a three-step method. First, we used a generalised Lomb–Scargle periodogram to obtain an initial estimate of the orbital period. This estimate was then used to perform a sinusoidal fit to the radial-velocity measurements. From this fit, we obtained initial estimates of the radial-velocity offset, semi-amplitude, and orbital phase. Finally, the initial guesses for all orbital parameters were used as starting points for a final Keplerian fit using a Markov Chain Monte Carlo (MCMC) technique to estimate the uncertainties of the fitted parameters, a commonly used approach in radial velocity studies \citep{BalanLahav2009}. We used SpinOS \citep{2021ascl.soft02001F} to perform this fit, which returned the fitted parameters (orbital period, phase, and eccentricity), with the 68\% confidence interval as corresponding uncertainties. We find an eccentricity $e=0$, indicating that any nonzero eccentricity is below the detection threshold of our MCMC fit. This aligns with our expectation for short-period binaries that are predicted to circularise on short timescales ($\sim10^7$ - $10^8$ years) due to tidal forces \citep{goldman1991orbital}. The resulting fit has a reduced chi-square ($\chi_\nu^2$) of 1.02, which indicates excellent agreement between the model and the observed data. The final phase-folded result is shown in the bottom panel of Figure \ref{fig:general}.


We find an orbital period of $P=3.601\pm0.002$ days, a time of conjunction of $T_0=2460955.471\pm0.004$ JD, and semi-amplitudes of $K_1=73.7\pm1.0$ km/s and $K_2=83.8\pm2.4$ km/s for the primary and secondary components, respectively.


\subsection{Source classification} \label{subsec:specclass}

In addition to determining the orbital parameters of the TYC 2834-1385-1 system, we also attempt a preliminary spectral classification based on its observed spectrum. As this is the first optical spectrum obtained for the source, no previous spectroscopic classification is available. Both stellar components exhibit very similar line strengths and absorption-line patterns, suggesting comparable atmospheric properties and spectral types. Together with the near-unity mass ratio, $q=M_2/M_1=K_1/K_2=0.88\pm0.03$, this suggests that TYC 2834-1385-1 consists of two similar stars.

Additionally, the spectrum of TYC 2834-1385-1 closely resembles that of EZ Peg, an RS Canum Venaticorum (RS CVn) system observed with HERMES during the same observing window, allowing for a direct comparison between the two sources. The density of metal absorption lines is comparable in both systems, suggesting that both sources have similar effective temperatures. However, the spectrum of TYC 2834-1385-1 exhibits systematically broader absorption lines, implying that its components are main-sequence stars, unlike the subgiant components of EZ Peg \citep{howell1986ez}. The spectrum of TYC 2834-1385-1 also shows strong double-lined calcium H and K emission, see Figure \ref{fig:general}. Such double-lined Ca II H and K emission lines are not characteristic of an RS CVn system \citep{1976ASSL...60..287H}, but are more characteristic of a BY Dra variable.



Finally, we determine a preliminary spectral classification for TYC 2834-1385-1. Several spectral features support a late-G to early-K classification. The prominent Ca II H and K lines, CH absorption around 4300 Å, and the strong Na I D doublet are consistent with late-type stars \citep{fernandez1993h,CHBandStars,NaDCitation}. The weaker Balmer lines relative to EZ Peg indicate a lower effective temperature, while the absence of prominent TiO bands rules out an M-type classification \citep{spectralclassificationbook,TiO_in_M}. Taken together, these features suggest a preliminary K0V classification for TYC 2834-1385-1.



\section{Discussion} \label{sec:discussion}

The results presented here show the first obtained optical spectrum of TYC 2834-1385-1 and establish it as a short-period, double-lined spectroscopic binary consisting of two similar late-type main-sequence stars. The preliminary K0V classification is consistent with the findings of \cite{Callingham2023}, who describe TYC 2834-1385-1 as “consistent with an early K-dwarf”. The strong double-lined Ca II H and K emission and late-type dwarf classification are consistent with TYC 2834-1385-1 being a BY Dra-type active binary. Further spectroscopic observations would allow the classification and individual properties of the two components to be refined, while photometric observations could constrain the system inclination and component masses.


\begin{acknowledgements}


Based on observations made with the Mercator Telescope, operated on the island of La Palma by the Flemish Community, at the Spanish Observatorio del Roque de los Muchachos of the Instituto de Astrofísica de Canarias.

Based on observations obtained with the HERMES spectrograph, which is supported by the Research Foundation - Flanders (FWO), Belgium, the Research Council of KU Leuven, Belgium, the Fonds National de la Recherche Scientifique (F.R.S.-FNRS), Belgium, the Royal Observatory of Belgium, the Observatoire de Genève, Switzerland and the Thüringer Landessternwarte Tautenburg, Germany.

Additionally, we thank Dr. A. Tkachenko for assistance with the least-squares deconvolution analysis; Dr. S. Rosu for obtaining additional spectra of TYC 2834-1385-1; Dr. J. Bodensteiner for assistance with the MCMC radial-velocity fitting and SpinOS; and M. Stroet, V. Anilkumar, and N. Swinkels for providing the spectrum-normalization code.

\end{acknowledgements}

\begin{contribution}

Alexander Hoogerbrug and Tamara A. S. Bleeker were responsible for writing and submitting the manuscript.
Elise Koo and Rudy Wijnands provided guidance during the observations and edited the manuscript.
Joseph R. Callingham assisted with the submission process and edited the manuscript.


\end{contribution}

%
\facilities{Mercator Telescope (HERMES)}

\software{LSD Denoiser \citep{2013A&A...560A..37T},  
          SpinOS \citep{2021ascl.soft02001F}, 
          Spectrum Normalizer \citep{python4esac_specnorm}
          }


\bibliography{sample701}{}
\bibliographystyle{aasjournalv7}



\end{document}